\documentclass[prl,showpacs,twocolumn,superscriptaddress,nofootinbib]{revtex4-1}

\usepackage{tikz}
\usepackage{pslatex}
\usepackage[T1]{fontenc}
\usepackage[latin1]{inputenc}
\usepackage{amssymb, amsmath, amsfonts, bm, amsthm, braket, mathtools}
\newcommand{\Eref}[1]{Eq.~(\ref{#1})}
\usepackage{multirow}
\usepackage{graphicx}
\usepackage{tabularx}
\usepackage{chngpage}
\input{epsf}
\usepackage[english]{babel}
\newcommand{\uw}{{\mu w}}
\newcommand{\lae}{\lower 2pt \hbox{$\, \buildrel {\scriptstyle <}\over {\scriptstyle\sim}\,$}}
\newcommand{\mathds}[1]{\text{\usefont{U}{dsrom}{m}{n}#1}}

\newcommand{\nsum}{\displaystyle\sum_{n=1}^N}
\newcommand{\ignore}[1]{}

\newtheorem{theorem}{Theorem}

\DeclarePairedDelimiterX{\norm}[1]{\lVert}{\rVert}{#1}

\usepackage{etoolbox}

\begin{document}
\title{Arbitrary Dicke-State Control of Symmetric Rydberg Ensembles}

\author{Tyler Keating}
\affiliation{Center for Quantum Information and Control (CQuIC), University of New Mexico, Albuquerque NM 87131}
\affiliation{Department of Physics and Astronomy, University of New Mexico,
Albuquerque NM 87131}

\author{Charles H. Baldwin}
\affiliation{Center for Quantum Information and Control (CQuIC), University of New Mexico, Albuquerque NM 87131}
\affiliation{Department of Physics and Astronomy, University of New Mexico, Albuquerque NM 87131}

\author{Yuan-Yu Jau}
\affiliation{Center for Quantum Information and Control (CQuIC), University of New Mexico, Albuquerque NM 87131}
\affiliation{Sandia National Laboratories, Albuquerque, NM 87185}

\author{Jongmin Lee}
\affiliation{Sandia National Laboratories, Albuquerque, NM 87185}

\author{Grant W. Biedermann}
\affiliation{Center for Quantum Information and Control (CQuIC), University of New Mexico, Albuquerque NM 87131}
\affiliation{Sandia National Laboratories, Albuquerque, NM 87185}

\author{Ivan H. Deutsch}
\affiliation{Center for Quantum Information and Control (CQuIC), University of New Mexico, Albuquerque NM 87131}
\affiliation{Department of Physics and Astronomy, University of New Mexico, Albuquerque NM 87131}

\begin{abstract}
We study the production of arbitrary superpositions of Dicke states via optimal control. We show that $N$ atomic hyperfine qubits, interacting symmetrically via the Rydberg blockade, are well-described by the Jaynes-Cummings Hamiltonian and fully controllable by phase-modulated microwaves driving Rydberg-dressed states. With currently feasible parameters, it is possible to generate states of $\sim 10$ hyperfine qubits in $\sim 1$ $\mu$s, assuming fast microwave phase switching time.  The same control can be achieved with a ``dressed-ground control'' scheme, which reduces the demands for fast phase switching at the expense of increased total control time.  
\end{abstract}
\pacs{34.50.-s,34.10.+x}
\maketitle

Creating entangled many-body states is a central challenge of quantum information science. Beyond their intrinsic interest as highly nonclassical states, such states are a resource for information processing protocols, including measurement-based quantum computation~\cite{measurement_computation},  error correction~\cite{magic_states}, and metrology beyond the standard quantum limit~\cite{metrology_lloyd}. In neutral atoms, one powerful tool for generating such states is the \emph{Rydberg blockade}, where the electric dipole-dipole interaction (EDDI) between high-lying Rydberg states suppresses excitation of multiple Rydberg states at a time~\cite{jaksch2000, lukin2001}. This effect has been used to entangle pairs of trapped atoms~\cite{saffmannature, browaeys, sandia_nature} and mesoscopic ensembles of atoms~\cite{saffman_ensembles, Kuzmich2012}.

Here, we apply the Rydberg blockade to many-body quantum state control.  Specifically, we study \emph{symmetric} ensemble control, in which one produces a target entangled state by applying a Hamiltonian that acts on every atom in the ensemble equivalently.  For an ensemble of $N$ qubits, this corresponds to controlling a Hilbert space spanned by the Dicke states, the symmetric subspace of $N$ spin-1/2 particles, with total spin $J=N/2$.  Control and measurement of Dicke states is more tractable since the symmetric subspace grows linearly with the number of particles, whereas in a general tensor-product space, the dimension grows exponentially.  This should allow us to develop new tools for control and measurement of many-body systems.  Dicke-state control has been demonstrated in ionic~\cite{dicke_ions} and photonic~\cite{dicke_photons} systems, and proposed for Bose-Einstein condensates~\cite{dicke_bec}.  A simple case of two-atom  symmetric control based on the Rydberg blockade was demonstrated by Jau {\em et al.} to produce Bell states~\cite{sandia_nature}.

\begin{figure} \label{fig:many_levels}
\includegraphics[scale=0.5]{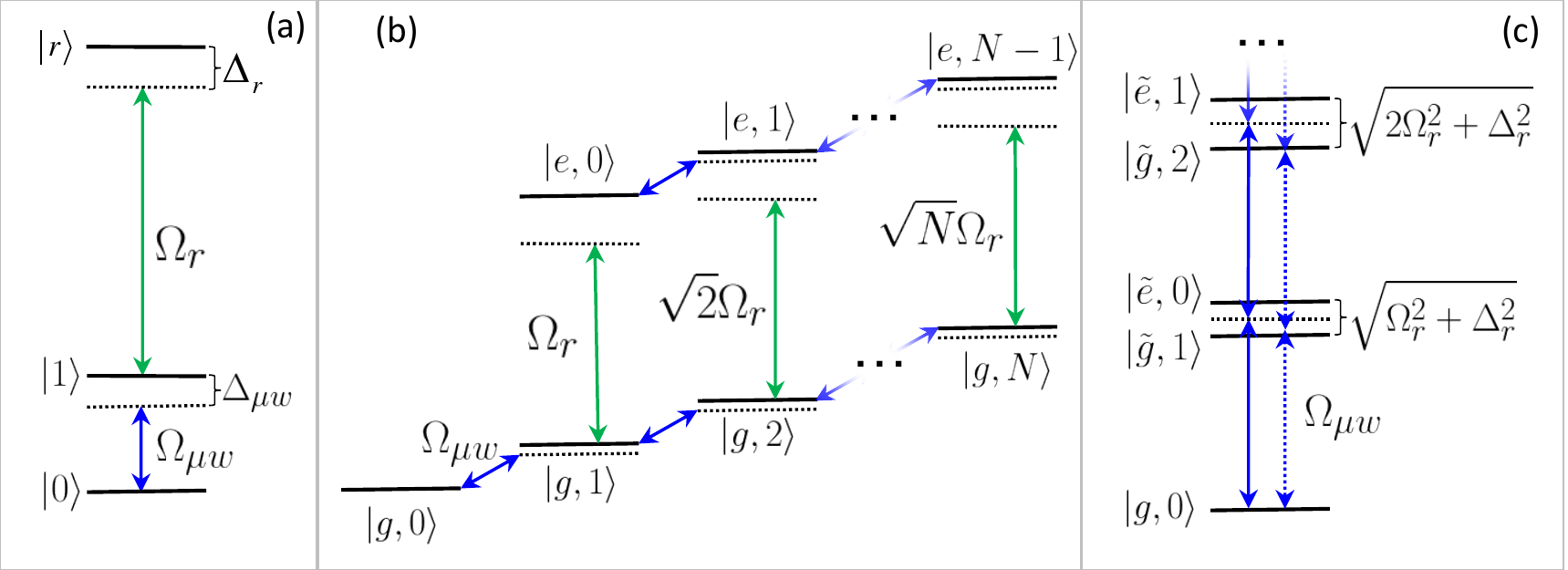}
\caption{(a) Basic level structure for the three-level atom: a qubit is encoded in the ground hyperfine states, and logical-$\ket{1}$  is optically coupled to a Rydberg state, while logical-$\ket{0}$ is far off resonance and effectively uncoupled. (b) Bare states for $N$ atoms, symmetrically coupled, under a perfect blockade. (c) $N$-atom dressed states, exhibiting the nonlinear JC ladder energy-level structure. Full Hilbert space control performs best when the microwave is tuned near resonance with the bare $\ket{0}\leftrightarrow\ket{1}$ transition (solid arrows), while dressed-ground control performs best when the microwave is tuned near the dressed-ground state transitions (dotted arrows).}
\end{figure}

To achieve Dicke-state control we will employ an isomorphism between the dynamics of the Rydberg-blockaded symmetric atomic ensemble, and the Jaynes-Cummings Model (JCM)~\cite{JCM, unanyan_2002, molmer_jc_2008, molmer_jc_2012, beterov_jc_2014}.  The nonlinear dynamics of the JCM have been well-studied in cavity QED and provide a powerful platform for quantum control~\cite{collapse_revival1, law_eberly, ion_jc, supercond_jc}.  To see this isomorphism, we consider a collection of $N$ atoms individually held in an array of optical dipole traps~\cite{Browaeys2016}.  For concreteness, we consider $^{133}$Cs atoms as employed in ~\cite{Hankin2014, sandia_nature, Lee2016} and encode qubits in the clock states, $\ket{0}\equiv\ket{6S_{1/2}, F=3,M_F=0}$ and \hbox{$\ket{1} \equiv \ket{6S_{1/2}; F=4,M_F=0}$} separated by hyperfine energy $\hbar \omega_{HF}$. We assume that the ensemble is uniformly illuminated by a 318 nm laser, coupling $\ket{1}$ to $\ket{r}\equiv\ket{n P_{3/2}, M_J=3/2}$ in every atom with the same Rabi frequency $\Omega_r$ and detuning $\Delta_r$ (Fig.~1).  Insofar as interactions are independent of the atoms' spatial positions, in second-quantization, and in the rotating fame at the laser frequency, the many-body Hamiltonian is
\begin{eqnarray} \label{eq:almost_JC}
 {H}=&E_0  {a}_0^{\dag} {a}_0 + (E_0+\hbar \omega_{HF})  {a}_1^{\dag} {a}_1+(E_0+\hbar \omega_{HF}-\hbar \Delta_r) {a}_r^{\dag} {a}_r \nonumber \\
&+\frac{\hbar \Omega_r}{2}( {a}_r^{\dag} {a}_1+ {a}_1^{\dag} {a}_r)+ {V}_{dd},
\end{eqnarray}
where $ {a}_i^\dag$ creates an atom in the state $\ket{i}$ symmetrically across the ensemble, so $[ {a}_i^\dag,  {a}_j ]= \delta_{i,j}$, the Bose commutation relations. When the EDDI, ${V}_{dd}$, is sufficiently strong across the whole ensemble, the analog of the Pauli exclusion principle allows only one Rydberg atom at a time, enforcing a perfect blockade.  For example, for $n=84$, with van der Waals coefficient $C_6/h = - 610$ GHz $\mu$m$^{6}$~\cite{Hankin2014}, and for $\Omega_r /2 \pi= 5$ MHz, the blockade radius is $r_B \equiv (C_6/\hbar \Omega_r)^{1/6} \approx  7.04$ $\mu$m. A $3\times 3$ square array of 9 atoms in dipole traps separated by 2 $\mu m$ are safely blockaded.  For these atoms, the dipole blockade thus effectively fermionizes the Rydberg state, and its creation operator now obeys an anti-commutation relation: $ {a}_r \rightarrow  {c}_r$, $\{ {c}_r^\dag,  {c}_r \} = 1$.  Making the Jordan-Wigner transformation from fermionic to Pauli operators, since we have one ``mode'', $ {c}_r\rightarrow {\sigma}_-$.  Taking $E_0=0$ and rewriting \Eref{eq:almost_JC} in terms of these operators gives
\begin{equation} \label{eq:really_almost_JC}
 H={H}_{JC}=\hbar \omega_{HF} {a}_1^{\dag} {a}_1+\hbar \omega_0 {\sigma}_+  {\sigma}_- +\hbar g ( {\sigma}_+ {a}_1+ {a}_1^{\dag} {\sigma}_-), 
\end{equation}
where $\hbar \omega_0 = \hbar \omega_{HF}-\hbar \Delta_r$ and $g = \frac{\Omega_r}{2}$. The dynamics of the many-body state is described by the familiar JC Hamiltonian~\cite{JCM}. Here, the presence or absence of a Rydberg excitation plays the role of the two-level atom in a conventional cavity QED setting, and the number of atoms in $\ket{1}$ takes the place of photons as the system's bosonic degree of freedom.

Under this mapping, the bare states of the JCM are symmetric superpositions of $n$ atoms in $\ket{1}$, with the remaining $N-n$ atoms distributed between $\ket{0}$ and $\ket{r}$ as
\begin{eqnarray}
&\ket{g,n} \equiv a_1^{\dag n} a^{\dag N-n}_0\ket{0}=\left\{\ket{0}^{\otimes N-n} \ket{1}^{\otimes n}\right\}_{sym} \\
&\ket{e,n} \equiv   c_r^\dag a_1^{\dag n}a_0^{\dag N-n-1}\ket{0}=\left\{\ket{0}^{\otimes N-n-1} \ket{1}^{\otimes n}\ket{r} \right\}_{sym}.
\end{eqnarray}
We recognize these states also as Dicke states, or eigenstates of a collective spin, with $J=\frac{N}{2},\frac{N-1}{2}$ for the ground and excited manifolds respectively.
We denote the Rydberg-dressed states, $\{\ket{\tilde{g},n}, \ket{\tilde{e},n-1}\}$, with energies $E_{\pm,n} = - \frac{\hbar \Delta_r}{2} \pm  \frac{\hbar}{2}{\rm sign}(\Delta_r)\sqrt{n \Omega_r^2 + \Delta_r^2}$ that represents the well-known JC ladder. 

We quantify the entangling power of the JC Hamiltonian by the nonlinear shift of the dressed states,
\begin{equation} \label{eq:kappa_form}
\kappa=\bra{\tilde{g},2}H_{JC}\ket{\tilde{g},2}-2\bra{\tilde{g},1}H_{JC}\ket{\tilde{g},1}.
\end{equation}
In the weak dressing limit, $|\Delta_r| \gg \Omega_r$, $\kappa\approx -\frac{\Omega^4}{8\Delta_r^3}$ and the nonlinearity of $H_{JC}$ is fully described by the two-body coupling $\kappa$ according to
\begin{equation} \label{eq:nkappa}
\bra{\tilde{g},n}H_{JC}\ket{\tilde{g},n}-n\bra{\tilde{g},1}H_{JC}\ket{\tilde{g},1}\approx(n^2-n)\frac{\kappa}{2}.
\end{equation}
For our atomic ensemble, on the ground manifold, the Hamiltonian is then a quadratic function of the collective spin,
\begin{equation}
H^{(g)}_{JC} = \frac{N}{2}\hbar\omega_{HF}+\left(\hbar\omega_{HF}+\frac{\hbar \Omega_r^2}{4 \Delta_r} +N \frac{\hbar \kappa}{2}\right) J_z +\frac{\hbar \kappa}{2} J_z^2.  
\end{equation}
This describes the one-atom light-shift plus entangling two-atom interactions that yield a one-axis-twisting Hamiltonian~\cite{kitagawa_ueda}.  This Hamiltonian produces cat states when applied to a spin coherent state, $e^{-i\kappa T J^2_z/2} (\ket{0}+\ket{1})^{\otimes N} =  e^{-i\pi/4}(\ket{0}^{\otimes N}+ i \ket{1}^{\otimes N})$, when $T=\pi/\kappa$~\cite{greiner_cat,  blatt_cat}.

We endeavor to go beyond cat states, employing quantum control to generate arbitrary target states in the ground Dicke subspace, $\ket{\Psi_{target}} =\sum_{n=0}^N c_{n} \ket{g,n}$.  In the context of cavity QED, such states correspond to the atom in $\ket{g}$ and a nonclassical state of the field, with up to $N$ photons.  The nonlinearity of the JCM provides numerous handles for achieving this with various degrees of control~\cite{collapse_revival1,law_eberly, ion_jc, supercond_jc}, 
Here, we show that the tools of optimal control can be used to generate fast state-to-state maps producing arbitrary target states.

In the setting of optimal control we consider a time-dependent Hamiltonian of the form $ {H}(t) =  {H}_{JC} +  {H}_c\left[\phi(t)\right]$, where the control Hamiltonian $ {H}_c$ is a functional of the waveform $\phi(t)$.  Through standard techniques, one can determine if the system is {\em controllable} on a Hilbert space of dimension $d$, meaning that there exists a $\phi(t)$ such that $ {H}(t)$ can generate any unitary map in the group SU($d$) after some time $T$. For our system, the total Hilbert space is $\mathcal{H} = \mathcal{H}^{(g)}_{J=N/2} \oplus \mathcal{H}^{(e)}_{J=(N-1)/2}$, corresponding to the ground/excited manifolds of the JCM with up to $N$ excitations. 

The Hilbert space and control Hamiltonians for our system bear close resemblance to the control of magnetic sublevels of hyperfine spins in ground-state alkali atoms, as employed in the seminal experiments of Jessen~\cite{jessen_control13, jessen_control}.  There, the combination of phase-modulated Larmor precession that generates SU(2) control on the spins with pairwise couplings between the sublevels of the two manifolds is sufficient for arbitrary control~\cite{merkel2008}.  Taking a similar strategy here, the couplings between the two manifolds are achieved by the Rydberg laser.  Arbitrary SU(2) control on each of the ground and excited manifolds corresponds to driving the system's bosonic degree of freedom in the JCM.  We can achieve this because, unlike a true harmonic oscillator, the atomic system is finite dimensional. Our control Hamiltonian is thus a microwave (or two-photon Raman) coupling $\ket{0}$ to $\ket{1}$ in each atom. The Rabi frequency and detuning are fixed at $\Omega_{\mu w}$ and $\Delta_{\mu w}=\omega_{\mu w}- \omega_{HF} $, respectively, but the microwave's phase can vary as a function of time. Assuming the microwave illuminates the entire ensemble symmetrically, in the frame rotating at the microwave frequency, the control Hamiltonian is
\begin{equation} \label{eq:Huw}
 {H}_c(t)=\frac{\hbar \Omega_{\mu w}}{2}(\cos\phi(t) {J}_x +\sin\phi(t)  {J}_y)-\hbar\omega_{\mu w}  ({J}_z +\frac{N}{2}),
\end{equation}
where $\phi(t)$ is the time-dependent phase. $H_c(t)$  generates arbitrary SU(2) rotations of the ground and excited manifolds.   Analogous to ~\cite{merkel2008}, $H(t)=H_{JC} + H_c(t)$ renders the system fully controllable, i.e., one can generate an arbitrary unitary map on the full  Hilbert space (see supplemental material).

Insofar as our system is controllable, we know there is always some (nonunique) waveform $\phi(t)$ that will generate any $\ket{\Psi_{target}}$ in the Dicke subspace from an initial fiducial state.  We consider here control waveforms  consisting of sequences of $s$ ``phase steps'' of length $\Delta t$, for a total run time of $T=s\Delta t$, 
as in~\cite{jessen_control13}. The range of possible control waveforms can be parameterized by an $s$-dimensional vector, $\vec{\phi}$.  We take as our fiducial state  $\ket{\Psi_0}=\ket{g,0}$. Turning on the Rydberg laser dresses the remaining eigenstates in the JC ladder, and control is performed in the dressed basis.  The target state of the control is thus $\ket{\widetilde{\Psi}_{target}} = \sum_n c_n \ket{\tilde{g},n}$.  After the control sequence, one would adiabatically undress the atom to achieve the target state in the bare ground Dicke subspace.

\begin{figure}
{\includegraphics[scale=0.4]{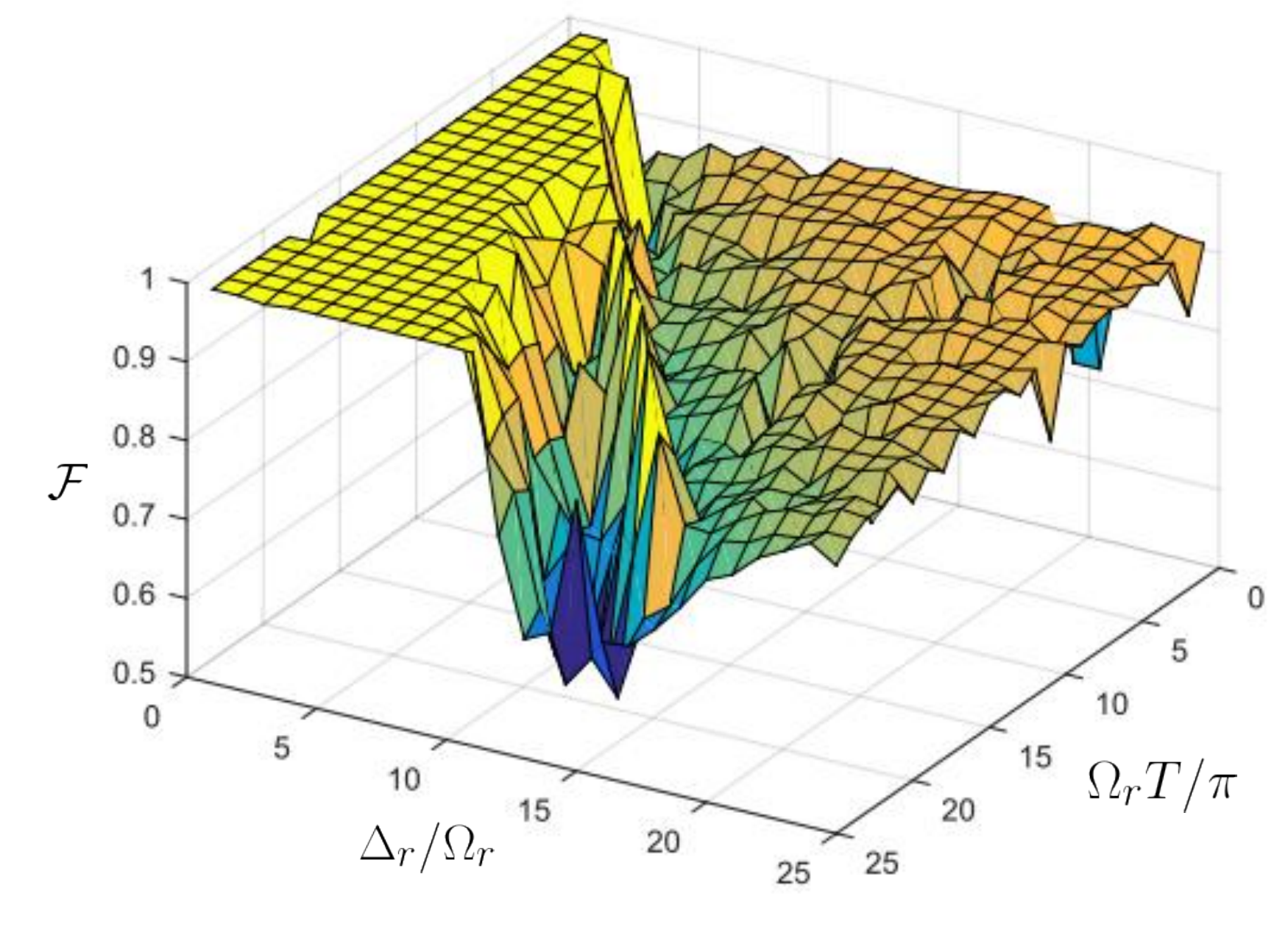}}
\caption{Simulated control fidelities to produce a six-atom cat state in the dressed basis, starting from $\ket{g,0}$, as a function of Rydberg laser detuning $\Delta_r$ and run time $T$, using $s=25$ phase steps.  For any $\Delta_r$, there is a minimum control time above which fidelity is arbitrarily close to one. As $\Delta_r$ increases, $\kappa$ decreases and the minimum control time gets longer.  Infidelities shown here are due solely to the quantum speed limit; decoherence is not included in the simulations.}
\label{fig:fidelity_plateau}
\end{figure}

 We seek a $\vec{\phi}$ such that the fidelity of the output with the target state, $\mathcal{F}(\vec{\phi})=|\bra{\tilde{\Psi}_{target}}U(\vec{\phi},T)\ket{\Psi_0}|^2$, is sufficiently high.  We find $\vec{\phi}$ with the well-known GRAPE gradient ascent algorithm~\cite{grape}.   The results are illustrated in Fig.~\ref{fig:fidelity_plateau} for a six-atom ensemble. The choice of optimal parameters such as laser/microwave power and detuning will depend on fundamental sources of error such as decoherence as well as practical experimental concerns. In particular, it is desirable to minimize the runtime and complexity of our protocol, so we typically seek the minimum $T$ and $s$ needed for high-fidelity control. It takes $2d-2$ real numbers to specify a $d$-dimensional target pure state, which puts a lower bound on $s$. For $N$ atoms including both ground and Rydberg symmetric states, this gives $s\geq 4N$. In practice, we find that this inequality can often be saturated. More heuristically, we can predict that the ``quantum speed limit'' is set by $ T\gtrsim\pi/\kappa$, the minimum time required to generate a cat state from separable state based on the one-axis twisting Hamiltonian, as discussed above. Whether this bound can be saturated depends on the choice of experimental parameters.

To achieve optimal fidelities, we wish to perform control in the shortest possible time compared with our system's decoherence time. Decoherence due to photon scattering, occurring at rate $\gamma$, is of particular concern, so maximizing $\kappa/\gamma$ is an important goal. Since $\kappa$ scales as $\Omega_r^4/\Delta_r^3$ in the weak dressing regime, it is highly sensitive to the power and detuning of the Rydberg laser. By contrast, $\gamma$ scales as $\Omega_r^2/\Delta_r^2$, so $\kappa/\gamma\propto\Omega_r^2/\Delta_r$ increases with increased laser power and decreased detuning. Based on this, increasing our dressing strength is a winning strategy in the fight against decoherence, and has the added benefit of reducing the total run time. This suggests that maximum laser power, at or near resonance, is the best choice of parameters.

No matter how short the control time is in principle, we must still have $s$ phase steps, and quickly switching a microwave's phase is not trivial. With resonant laser power that yields a Rabi frequency of a few MHz and $\sim10$ atoms, the required $\Delta t$ per phase step can easily shrink to tens of nanoseconds or less. Demands on the microwave switch time are even more strict, since the phase must change quickly enough to preserve the piecewise-constant approximation of $\phi(t)$. The number of steps in the control waveform, then, is a primary limiting factor in the speed and feasibility of this protocol. Once the  control time is limited by experimental restrictions on $\Delta t$ rather than by $\kappa$, increasing $\kappa$ is no longer beneficial; stronger dressing will only increase $\gamma$ and other sources of error without any offsetting benefit of control speed. On the other hand, as long as $\kappa$ is the limiting factor, increased dressing strength is advantageous as per the reasoning above. The optimal parameter regime, therefore, is highly dependent on the particulars of the experiment: $\Delta_r$ should be large enough to make the two speed limits match, if possible, but no higher.

Since phase switching requirements limit the speed of our protocol, control could be significantly accelerated by reducing the number of phase steps needed. This can be achieved if we can adiabatically eliminate the dressed-excited states, and restrict the dynamics to the dressed-ground manifold, the Dicke subspace of interest.  In this case the dimension of the control Hilbert space is cut in half, and we need only $2N$ parameters to specify a target in it. We clearly see that this is possible in the far off resonance limit.  Then the JC Hamiltonian on the ground manifold takes the form of a quadratic light shift, \Eref{eq:nkappa}.  This, together with SU(2) control generated by the microwaves renders the qudit $J$ fully controllable on SU($2J+1$)=SU($N+1$). 

More generally, we return to \Eref{eq:Huw}, describing the effect of coupling induced by the microwave or Raman transition. In the bare basis the microwave couples $\ket{0}$ to $\ket{1}$ without acting on $\ket{r}$, so $\bra{e,m}H_{\mu w}\ket{g,n}=0$. By contrast, dressed Rydberg states $\ket{\tilde{e},n}$ have some $\ket{g,n}$ character, so the microwave coupling between dressed-ground and Rydberg states is nonzero. In the weak dressing regime we can approximate the magnitude of this coupling as $\bra{\tilde{e},m}H_{\mu w}\ket{\tilde{g},n}\approx \sin\frac{\theta_n}{2}\bra{g,m}H_{\mu w}\ket{g,n} \approx \frac{\sqrt{n}\Omega_r}{2|\Delta_r|}\bra{g,m}H_{\mu w}\ket{g,n}$.  The effective Rabi rate is suppressed by an order of $\Omega_r/\Delta_r$ for dressed ground-excited coupling compared to dressed ground-ground couplings. Excitation is also suppressed by microwave detuning. The saturation parameter is on the order of $\Omega_{\mu w}^2/\Delta_r^2$. Combining these suppressing factors, we find that the microwave will approximately preserve dressed-ground population as long as ${\sqrt{N}\Omega_r\Omega_{\mu w}^2}/{\Delta_r^3}\ll 1$.  Under this condition, we find that dressed-ground control can be performed in $2N$ phase steps, as expected. Because this condition requires a large $\Delta_r$, it goes hand in hand with a small $\kappa$, so ground manifold control is much slower than full Hilbert space control if phase steps are allowed to be arbitrarily short. Whether the tradeoff between $\kappa$ and $s$ is worthwhile will depend on the $N$ and the minimum $\Delta t$ in a given experiment.

\begin{figure}
\includegraphics[scale=0.4]{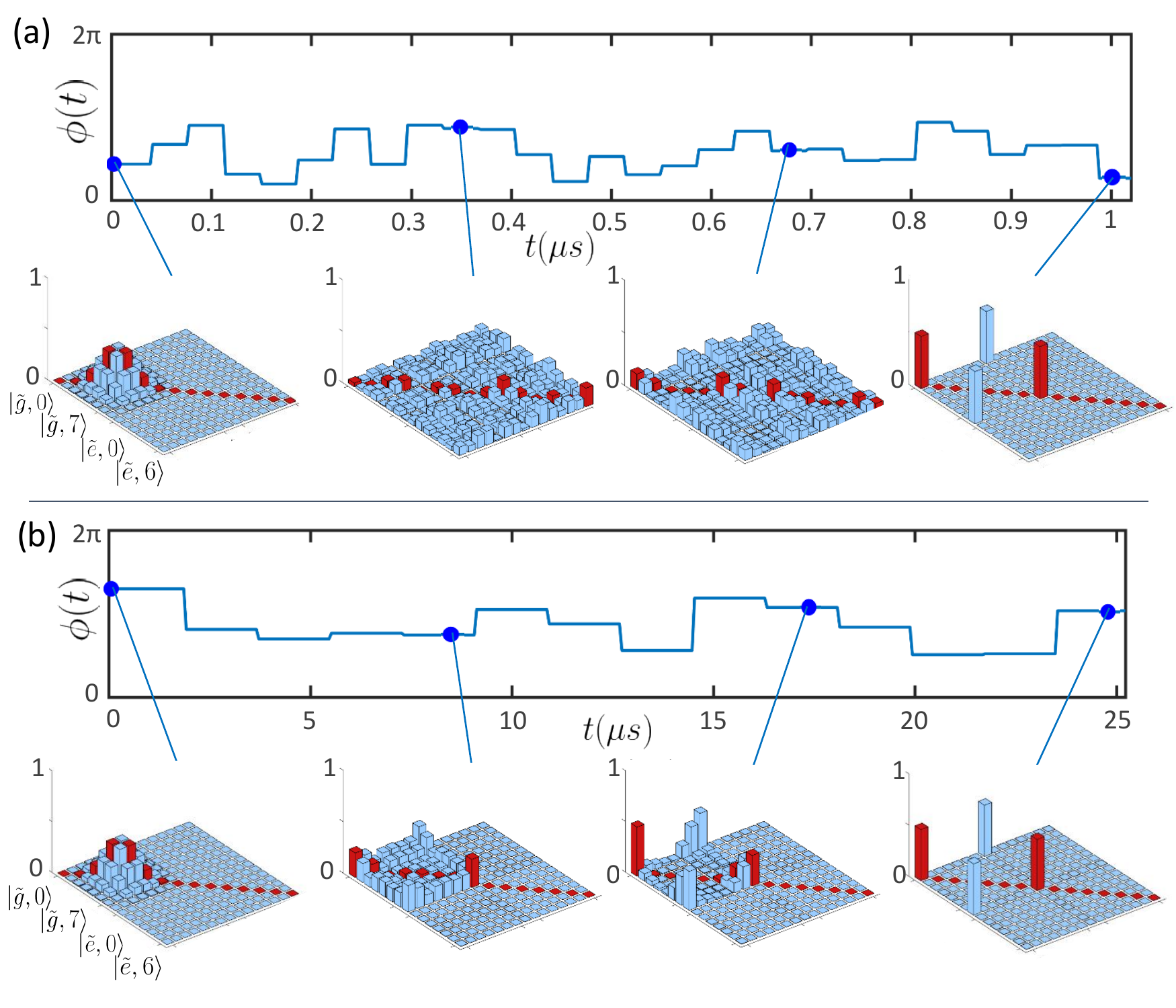}
\caption{Mapping the 7-atom spin coherent state $(\ket{0}+\ket{1})^{\otimes 7}/\sqrt{2^7}$ to the cat state $(\ket{0}^{\otimes 7} + \ket{1}^{\otimes 7})/\sqrt{2}$. Line plots show the microwave phase $\phi(t)$ found via optimal control, and bar charts show the real part of the $15\times15$ density matrix at various snapshots in time. (a) Full Hilbert space control: $\Omega_r/2\pi= 5$ MHz, $\Delta_r/2\pi=2.5$ MHz, $\Omega_{\mu w}/2\pi=12.5$ MHz, and $ \Delta_{\mu w}/2\pi  = 1.25$ MHz.  All 15 dressed states are traversed during control, so $4N=28$ phase steps are needed. (b) Dressed-ground control: $\Omega_r/2\pi= 5$ MHz, $\Delta_r/2\pi=15$ MHz, $\Omega_{\mu w}/2\pi=100$ kHz, and $\Delta_{\mu w}/2\pi= -400$ kHz. Population remains in the 8 dressed-ground states, so only $2N = 14$ phase steps are needed, but weaker Rydberg coupling reduces $\kappa$ by more than an order of magnitude with a commensurate increase in run time.}
\label{fig:evolution_comparison}
\end{figure}

Dressed-ground and full Hilbert space control are optimized with qualitatively different choices of microwave and laser parameters. When we control the whole Hilbert space, the system traverses both ground and excited states to get to its destination, so all states must be coupled strongly to each other. Since the light shift provides a gap between the ground and Rydberg manifolds and is of order $\sqrt{\Omega_r^2+\Delta_r^2}$, $\Omega_{\mu w}$ needs to be at least that large to strongly drive both transitions at once. Both to relax this condition and to maximize interaction strength, $\Delta_r$ should be kept small compared to $\Omega_r$. The microwave resonance should also be tuned approximately halfway between the ground and Rydberg states in the rotating frame ($\Delta_{\mu w} \approx\Delta_r/2$), so that all manifolds are roughly equally coupled (see Fig.~1c). If these conditions are not met, population transfer between the ground and Rydberg manifolds will be slow compared to intra-manifold transfer, and control can be bottlenecked by population getting ``stuck'' in the Rydberg manifold for extended periods.

On the other hand, dressed-ground control relies on the assumption of adiabatic elimination of the Rydberg manifold, so parameters should be chosen to \emph{minimize} coupling between ground and excited manifolds. $\Omega_{\mu w}$ should be small compared to the light shift gap, and a large $\Delta_r$ makes this easier to accomplish. Likewise, the microwave should be tuned near resonance with the transitions between dressed-ground states to allow strong dressed ground-ground coupling with minimal dressed ground-excited coupling. The microwave Rabi frequency also should be large compared $\kappa$ to ensure that off resonant driving to the excited dressed states is negligible over the entire control time.  Thus $\Omega_{\mu w}$ should scale inversely with $\Delta_r$. If these conditions are not met, significant population can leak into the dressed-excited manifold, where $2N$ free parameters are no longer enough to bring it back to the dressed-ground manifold. Dressed-ground and full Hilbert space control thus provide two complimentary methods that function well in different regimes. The two methods produce waveforms, each optimal for its respective parameter regime, that reach the same destination in Hilbert space but take qualitatively different paths to get there. This is illustrated in Fig.~\ref{fig:evolution_comparison}, which shows how both types of control can be used to produce a 7-atom cat state.  In both cases we employ the JCM, Eq.~(\ref{eq:really_almost_JC}), plus microwave control.  Fast full Hilbert-space control is achieved in 1 $\mu$s with phase steps of 35.7 ns; dressed-ground state control requires 25 $\mu$s but with phase steps of 1.79 $\mu$s.

Finally, all of the analysis in this work assumed a perfectly known model Hamiltonian and neglected decoherence and experimental noise.  Of particular importance, our JCM requires a perfect blockade.  For the parameters considered here, this limits us to consider ensembles with $\sim 10$ atoms, though employing Rydberg states at higher principle quantum numbers, or by packing atoms closer together in an optical lattice, one can reach $\sim 100$ atoms with a perfect blockade as seen in recent experiments~\cite{Bloch2015}. Fundamental decoherence occurs via spontaneous decay of the Rydberg state.  While the atomic lifetimes are long compared to the control times considered here, experiments show shorter coherence times that are still unexplained~\cite{Bloch2015}.  Reducing this decoherence will be essential to achieve high fidelity control.  Technical noise includes sensitivity to background electric fields, which can be managed with a proper experimental approach~\cite{Hankin2014, Lee2016}.  Other technical challenges include uncertainties in the parameters of the Hamiltonian including Rabi frequencies and detunings.  These may be mitigated with the techniques of {\em robust control}, which has been an essential tool to achieve high fidelity control of qudits~\cite{jessen_control}. 

This work was supported by the Laboratory Directed Research and Development program at Sandia National Laboratories and the Center for Quantum Information and Control (CQuIC), under NSF Grant No. PHY-1521016. This material is based upon work supported by the National Science Foundation under Grant No. PHY-1606989

\newpage
\section{Supplemental Material --- Proof: $H_{JC}$ With Microwaves is Controllable} \label{apx:control}

The total Hamiltonian for a general control task consists of a constant part $H_0$ and one or more adjustable parts $H_j$,
\begin{equation}
H(t)=H_0+\displaystyle\sum_j c_j(t)H_j,
\end{equation}
where $c_j(t)$ are the time-dependent control parameters. A $d$-dimensional system described by such a Hamiltonian is controllable if and only if the operators $\{H_0,H_1,...,H_n\}$ are a generating set for the Lie algebra $\mathfrak{su}(d)$. Therefore, we can show that a system is controllable by generating all elements of $\mathfrak{su}(d)$ through nested commutators and linear combinations of $\{H_0,H_1,...,H_n\}$. In the case we consider here, our full Hamiltonian is a combination of the Jaynes-Cummings and microwave-control Hamiltonians (Eqs.~2 and 9) from the main text: $H(t)=H_{JC}+H_c(t)$. Splitting this into constant and variable components gives
\begin{equation}
\begin{split}
H_0=&-\Delta_\uw(J_z+\frac{N}{2})+(E_{HF}-\Delta_r)P_e\\
&+\displaystyle\sum_{n=1}^{N}\frac{\sqrt{n}\Omega_r}{2}(\ket{g,n}\bra{e,n-1}+\ket{e,n-1}\bra{g,n}),\\
H_1=&J_x,\quad c_1(t)=\Omega_\uw\cos(\phi(t)),\\
H_2=&J_y,\quad c_2(t)=\Omega_\uw\sin(\phi(t)),
\end{split}
\end{equation}
where $P_{e(g)}=\sum_n\ket{e(g),n}\bra{e(g),n}$ is the projector onto the Rydberg (ground) manifold. Our goal is to produce the full Lie algebra for our system through commutators and linear combinations of these three Hamiltonians. Before beginning, we take three steps to simplify subsequent notation. First, we define the ``coupling'' portion of $H_0$,
\begin{equation}
H_C\equiv\displaystyle\sum_{n=1}^{N}\frac{\sqrt{n}\Omega_r}{2}(\ket{g,n}\bra{e,n-1}+\ket{e,n-1}\bra{g,n}).
\end{equation}
Second, without loss of generality, we switch to units of energy with $\Omega_r/2=1$, leaving
\begin{equation}
H_0\rightarrow H_C-\Delta_\uw(J_z+\frac{N}{2})+(E_{HF}-\Delta_r)P_e.
\end{equation}
Third, because the trace of $H_0$ will at most contribute an overall phase which can be ignored, we subtract $-\Delta_\uw\frac{N}{2}+\frac{E_{HF}-\Delta_r}{2}$ from the overall energy to leave $H_0$ traceless:
\begin{equation}
H_0\rightarrow H_C-\Delta_\uw J_z+\frac{E_{HF}-\Delta_r}{2}(P_e-P_g).
\end{equation}
Note that $H_1$ and $H_2$ are entirely off-diagonal operators, and so are already traceless.

Immediately, we can commute $J_x$ and $J_y$ to get
\begin{equation}
\begin{split}
J_z=\nsum &\Big(n\ket{g,n}\bra{g,n}+(n-1)\ket{e,n-1}\bra{e,n-1}\Big)\\
&-\frac{N}{2}P_g-\frac{N-1}{2}P_e\\
=H_C^2&-\frac{1}{2}P_e-\frac{N}{2}\mathds{1}.
\end{split}
\end{equation}
Because $J_z$ is diagonal in the $\{\ket{g,n},\ket{e,n}\}$ basis, it commutes with the projectors in $H_0$, and their commutator is greatly simplified,
\begin{equation}
\begin{split}
[H_0,J_z]&=[H_C,J_z]=[H_C,H_C^2-\frac{1}{2}P_e-\frac{N}{2}\mathds{1}]=-\frac{1}{2}[H_C,P_e]\\
&=\frac{1}{2}\nsum \sqrt{n}\Big(\ket{e,n-1}\bra{g,n}-\ket{g,n}\bra{e,n-1}\Big)\\
&\equiv \frac{i}{2} \bar{H}_C.
\end{split}
\end{equation}
Commuting with $J_z$ again,
\begin{equation}
\begin{split}
[\bar{H}_C,J_z]=&i\nsum \sqrt{n}\Big((n-1-\frac{N-1}{2})\ket{g,n}\bra{e,n-1}\\
&\hspace{.3cm}-(n-\frac{N}{2})\ket{e,n-1}\bra{g,n}-(n-\frac{N}{2})\ket{g,n}\bra{e,n-1}\\
&\hspace{.3cm}+(n-1-\frac{N-1}{2})\ket{e,n-1}\bra{g,n}\Big)\\
=&-\frac{1}{2}i\nsum\sqrt{n}\Big(\ket{g,n}\bra{e,n-1}+\ket{e,n-1}\bra{g,n}\Big)\\
&=-\frac{1}{2}iH_C.
\end{split}
\end{equation}
We can now use $H_C$ directly in subsequent steps, which will avoid complicating terms from the microwave and laser detunings. Our next step is to commute $H_C$ with $\bar{H}_C$, giving
\begin{equation}
\begin{split}
[H_C,\bar{H}_C]&=2i\nsum n\Big(\ket{e,n-1}\bra{e,n-1}-\ket{g,n}\bra{g,n}\Big)\\
&=2i\Big(P_e(J_z+\frac{N-1}{2}+1)P_e-P_g(J_z+\frac{N}{2})P_g\Big)\\
&\equiv 2i\bar{J}_z.
\end{split}
\end{equation}
This operator breaks the symmetry between the manifolds, and we can combine it with the original $J_z$ to get operators projected onto each manifold individually,
\begin{align}
&J_z - \bar{J}_z = 2P_gJ_z P_g + \frac{N}{2}P_g - \frac{N+1}{2}P_e,\\
&J_z+\bar{J}_z=2P_eJ_z P_e - \frac{N}{2}P_g + \frac{N+1}{2}P_e.
\end{align}
Note that the projectors commute with any operator that does not couple the two manifolds, so we can ignore them when commuting these projected $J_z$'s with other $J_i$'s. The resulting commutators allow us to spread the ground- and Rydberg-projections to $J_x$ and $J_y$,
\begin{align}
&[J_z-\bar{J}_z,J_j]=2\epsilon_{zjk}P_{g}J_k P_{g},\\
&[J_z+\bar{J}_z,J_j]=2\epsilon_{zjk}P_{e}J_k P_{e},
\end{align}
where $\epsilon_{zjk}$ is the Levi-Civita symbol with its first index fixed as $z$. This gives us independent SU(2) rotations of the two manifolds.

To get from SU(2) rotation to complete control, we invoke a theorem due to Seth Merkel~\cite{merkel_thesis}:
\begin{theorem}
Consider a manifold $M$ that is describable by a collective pesudo-spin. Let $T$ be an operator that has nonzero overlap with at least one irreducible, rank-two tensor operator on said spin. Then $M$ is controllable with the Hamiltonians $\{J_x,J_y,T\}$.
\end{theorem}
Based on this, we need to generate a rank-2 irreducible Hamiltonian for each manifold to make it controllable. This can be accomplished by commuting the projected $J_z$ with $\bar{H}_C$:
\begin{align}
\begin{split}
[\bar{H}_C,P_gJ_z P_g]&=-i\nsum n\sqrt{n}\Big(\ket{g,n}\bra{e,n-1}+\ket{e,n-1}\bra{g,n}\Big)\\
&\hspace{.3cm}+\frac{N}{2}i\nsum\sqrt{n}\Big(\ket{g,n}\bra{e,n-1}+\ket{e,n-1}\bra{g,n}\Big)\\
&\equiv-i H_C'+\frac{N}{2}iH_C
\end{split}\\
\begin{split}
[H_C',\bar{H}_C]&=2i\nsum n^2\Big(\ket{e,n-1}\bra{e,n-1}-\ket{g,n}\bra{g,n}\Big)\\
&=2i\Big(P_e (J_z+\frac{N-1}{2}+1)^2 P_e - P_g (J_z+\frac{N}{2})^2 P_g\Big)\\
&=2i\Big(P_eJ_z^2P_e-P_gJ_z^2P_g+(N-1)P_e J_z P_e\\
&\hspace{.3cm} - NP_g J_z P_g + \left(\frac{N+1}{2}\right)^2P_e - \left(\frac{N}{2}\right)^2P_g\Big).
\end{split}
\end{align}
This Hamiltonian has terms quadratic in $n$; we now condense it before commuting it with $P_gJ_xP_g$ to obtain a nonlinearity in the ground manifold alone. The third and fourth terms of the Hamiltonian are $J_z$ on the ground and Rydberg manifolds, respectively, which we have already generated and can subtract away. The first, fifth, and sixth terms, meanwhile, commute with $P_gJ_xP_g$, so we need only consider the second ($J_z^2$) term, which produces the commutator,
\begin{equation}
\begin{split}
[P_gJ_z^2P_g,P_gJ_x P_g]=iP_g\Big(J_z J_y + J_y  J_z\Big)P_g.
\end{split}
\end{equation}
This is an anti-commutator between two $J$'s, so it is an irreducible rank-2 operator. The same procedure with $P_eJ_zP_e$ gives a comparable operator for the Rydberg manifold, so we now have SU(2) rotations plus a rank-2 operator for both manifolds. This means that they are independently controllable.

All that remains is to join the two manifolds together, and to show that they are controllable as a whole as well as separately. For this, we invoke another theorem due to Seth Merkel~\cite{merkel_thesis}:
\begin{theorem}
Consider two subspaces, $L$ and $M$. Let $\ket{\ell}$ and $\ket{m}$ be particular states in each of these spaces, respectively. If $L$, $M$, and the subspace spanned by $\{\ket{\ell},\ket{m}\}$ are each independently controllable, then the full space $L\oplus M$ is controllable.
\end{theorem}
We have already shown the controllability of the two subspaces, so we just need to show controllability of any subspace consisting of one state from each. We arbitrarily choose $\{\ket{g,1},\ket{e,0}\}$. Since both manifolds are controllable, we can generate any traceless Hamiltonians within them. In particular, we can generate $\ket{g,0}\bra{g,0}-\ket{g,2}\bra{g,2}$ on the ground manifold and $\ket{e,0}\bra{e,0}-\ket{e,1}\bra{e,1}$ on the Rydberg manifold. Summing these and commuting with $H_C$ gives
\begin{equation}
\begin{split}
\Big[H_C,&\ket{g,0}\bra{g,0}-(\ket{g,2}\bra{g,2}+\ket{e,1}\bra{e,1})+\ket{e,0}\bra{e,0}\Big]\\
&=[H_C,\ket{e,0}\bra{e,0}]\\
&=2\Big(\ket{g,1}\bra{e,0}-\ket{e,0}\bra{g,1}\Big)=2i\sigma_y
\end{split}
\end{equation}
where $\sigma_i$ denotes a Pauli operator on the two-state subspace. Commuting this with the coupling Hamiltonian one last time,
\begin{equation}
[H_C,\sigma_y]=2i\Big(\ket{g,1}\bra{g,1}-\ket{e,0}\bra{e,0}\Big)=2i\sigma_z.
\end{equation}
Two Pauli operators give us control over the two-state subspace, and therefore over the entire space. $\square$

\bibliographystyle{apsrev4-1}
\bibliography{RydbergSymmetricControl_Arxiv_Final}

\end{document}